# Structural origin for the dipole x-ray resonant scattering in the low temperature phase of $Nd_{0.5}Sr_{0.5}MnO_3$ manganite

Javier Herrero-Martín, Joaquín García, Gloria Subías, Javier Blasco and Maria Concepción Sánchez

*Instituto de Ciencia de Materiales de Aragón, Consejo Superior de Investigaciones Científicas y Universidad de Zaragoza, Pza. San Francisco s/n, 50009 Zaragoza, Spain.*

Corresponding author: Joaquín García Ruiz

E-mail: jgr@unizar.es

Telf. 34-976-761225

FAX. 34-976-761229

Instituto de Ciencia de Materiales de Aragón- CSIC

Facultad de Ciencias, Universidad de Zaragoza, 50009 Zaragoza, Spain

**Abstract**

We have investigated the low temperature phase of a $Nd_{0.5}Sr_{0.5}MnO_3$ single crystal by x-ray resonant scattering at the Mn K-edge of the (3 0 0), (0 3 0) and (0 5/2 0) reflections. Strong resonances were observed for the $\sigma-\sigma$' channel in the (3 0 0) and (0 3 0) reflections and for the $\sigma-\pi$' channel in the (0 5/2 0) reflection. These resonances show a $\pi$-periodicity on the azimuthal angle, having the intensity at the minimum position almost zero. The intensity dependence on the photon energy, azimuthal angle and polarisation dependencies has been analysed using a semi-empirical structural model. Contrary to previous claims of charge ($Mn^{3+}$-$Mn^{4+}$) and orbital ordering in this compound, our results show that the dipole resonant superlattice reflections can be explained by the presence of two types of Mn sites with different local geometric structure. One of the Mn sites is surrounded by a tetragonal-distorted oxygen octahedron whereas the other site has a nearly regular octahedral environment. This model also establishes that no real space charge ordering is needed to explain the experimental data. Intermediate valence states according to a fractional charge-segregation $Mn^{+3.42}$-$Mn^{+3.58}$ were deduced.

## 1. INTRODUCTION

Manganese mixed valence perovskites of the type $RE_{1-x}A_xMnO_3$, $RE_{1-x}A_{1+x}MnO_4$ and $RE_{2-2x}A_{1+2x}Mn_2O_7$ (RE: rare-earth, A: alkaline-earth), hereafter referred as manganites, have attracted a great deal of attention during the last years by the unusual physical properties such as colossal magnetoresistence[1-3]. Depending on the formal valence state of the Mn atom they show a variety of magnetic and electrical phases, including ferromagnetic-metal, antiferromagnetic-insulators, and phases identified as charge-orbital ordered (COO)[2-4].

The general principle to understand this behavior was based on a single atom ionic picture, where (1) $Mn^{3+}$ and $Mn^{4+}$ ionic states are implicitly (temporal or spatial) distinguished and (2) the atomic 3d states are splitted by the octahedral crystal field into $t_{2g}$ and $e_g$ orbitals, the electronic configurations for $Mn^{3+}$ and $Mn^{4+}$ ions being the high-spin $t_{2g}^3e_g^1$ and $t_{2g}^3$, respectively. Within these premises, the formation of an ionic ordered sequence of $Mn^{3+}$ and $Mn^{4+}$ ions at low temperatures, as proposed years ago for $Fe^{2+}$/ $Fe^{3+}$ ordering at the Verwey transition in magnetite[5], is its natural consequence. In this way, the orbital ordering (OO) proposed for some manganites is also a consequence of the assumed ionic model. The occurrence of an electronic degenerated E state ( $t_{2g}^3e_g^1$ configuration of $Mn^{3+}$ ion) should induce a tetragonal distortion of the octahedron coupled simultaneously to the splitting of the $e_g$ states (Jahn-Teller theorem) thus producing a directional orientation of the lowest energy $e_g$ orbital[6]. Then, the OO proposed for $LaMnO_3$ and other related manganites is ascribed to the ordering of the tetragonal Jahn-Teller distortions.

Manganites near the half doping (i.e. $x\approx 0.5$), where the formal valence on the Mn atom is +3.5, are the object of special attention because charge ($Mn^{3+}$/ $Mn^{4+}$) and orbital ordering was proposed as the fingerprint of the CE-antiferromagnetic insulating low temperature phase. For instance, a charge-ordered (CO) phase was proposed at low temperatures for $Pr_{0.5}Ca_{0.5}MnO_3$, [7] $Nd_{0.5}Sr_{0.5}MnO_3$, [8] $La_{0.5}Ca_{0.5}MnO_3$,[9,10] $La_{0.5}Sr_{1.5}MnO_4$ [11,12] or $LaSr_2Mn_2O_7$ [13]. The classical model proposed to describe this low temperature phase is a one-dimensional zigzag chain of Mn atoms in the *ab* plane, coupled antiferromagnetically to each other (checkerboard pattern, fig. 1). This model has been supported by the structural determination given by Radaelli et al.[9,10] showing that two non-equivalent crystallographic sites are present in the low temperature phase of $La_{0.5}Ca_{0.5}MnO_3$ for the Mn atom. In terms of the single ion picture, one of the $MnO_6$ octahedron is tetragonally distorted and was ascribed to a Jahn-Teller $Mn^{3+}$ ion while the other one shows a regular octahedron of oxygen atoms and was assigned to the $Mn^{4+}$ ion. Moreover, the real image of stripes observed by electron microscopy was also considered as a proof of charge ordering[14,15]. Within this assumption, the structural anisotropy of the so-called $Mn^{3+}$ ion was identified as due to d-orbital occupancy giving rise to the term of orbital-ordering.

In spite of the widely acceptance of this model, some papers either theoretical or experimental criticise it. Most of the crystallographic determinations[16-19] essentially follow the pioneering work of Radaelli et al.[9]. All these studies proposed a monoclinic $P2_1/m$ symmetry with a doubled *b*-axis (*Pbnm* or *Ibmm* setting) for the low-temperature phase. Single crystal neutron diffraction experiments on $Pr_{0.6}Ca_{0.4}MnO_3$ instead, proposed an alternative structure. Daoud-Aladine et al.[20] described the low temperature phase of this sample as formed by pairs of manganese ions having a very similar oxygen octahedral environment. Accordingly, they concluded the absence of CO and proposed the formation of a Zener polaron as electronic localisation mechanism. Nevertheless, none of the structural models reported Mn-O interatomic distances that agree with those in pure $Mn^{3+}$ and $Mn^{4+}$ perovskites ($LaMnO_3$ and $CaMnO_3$, for example). So independently of the ordering pattern, a question remains: what is the limit to speak in terms of $Mn^{3+}$ and $Mn^{4+}$ ionic states? Despite this contrasting experimental evidence, recent theoretical calculations suggest a band-insulator picture for the insulating CO state instead of a single atom localisation mechanism[21-23]. Moreover, other authors proposed that the charge segregation mainly occurs on the oxygen atoms[24].

The COO phenomena have been recently studied by means of x-ray resonant scattering (XRS). This technique measures the intensity of reflections either forbidden by crystal symmetry or allowed with very low-intensity as a function of the incident photon energy across an absorption edge. For these reflections, the structure factor is given by the difference (roughly) of atomic scattering factors. This fact makes the Thomson scattering be zero or nearly zero and strong resonances coming from the anomalous scattering factor of these atoms can be observed at the absorption edge[25,26]. Note that the anomalous atomic scattering factor has a tensorial character, being a symmetric tensor of range 2 for virtual electronic dipolar transitions (the main contribution). The tensorial character implies that the three components of the diagonalized tensor and the orientation of the tensor in the crystal must be taken into account. The scattered intensity will depend on both, the incident beam polarisation and the azimuthal scattering angle. In this sense, those reflections arising from a different spatial orientation of the scattering tensor in the crystal are called "*Templeton or ATS (Anisotropic Tensor Susceptibility)*" reflections[25].

The claim for ionic $Mn^{3+}$-$Mn^{4+}$ charge-ordering in several half-doped manganites[27-32] was based on the observation of a strong resonance at the Mn K-edge for (0,odd,0) superlattice reflections in some 3-dimensional manganites as $Pr_{0.5}Ca_{0.5}MnO_3$ or $Nd_{0.5}Sr_{0.5}MnO_3$. On the other hand, the observation of an ATS resonance for the (0, odd/2, 0) forbidden reflections has been considered as the experimental proof of d-orbital ordering. The same interpretation was given for the corresponding (h/2, k/2, 0) and (h/4, k/4, 0) reflections in the 2-dimensional manganites $La_{0.5}Sr_{1.5}MnO_4$ and $LaSr_2Mn_2O_7$. This interpretation is still controversial from both, the experimental[33-37] and the theoretical[22,38-42] points of view: First, the analysis of the so-called CO reflections was made considering the anomalous scattering factor as scalar despite the azimuthal and polarisation behavior observed for these reflections. For instance, the low or zero intensity of the CO resonance at particular azimuthal angles was not considered. Second, the observation of ATS reflections was considered as a direct proof of d-orbital ordering. It is well-known that the anisotropy of the anomalous scattering factor mainly arises from the low symmetry of the local structure around the anomalous atom. In fact, the extended experience on the intimately related x-ray absorption spectroscopy (XAS) technique shows that anisotropy is observed for any kind of atoms with asymmetric local geometry, not only for non-filled d-metals[43]. In addition to these general criticisms, XAS experiments at the Mn K-edge of $RE_{1-x}Ca_xMnO_3$ ( RE = La, Ca) manganites have shown that the spectra can not be described by a mixture of $Mn^{3+}$ and $Mn^{4+}$ ionic states[44-46].

The experimental confirmation about the existence or absence of integer ionic states in mixed valence oxides is a matter of fundamental interest. The non-identification of integer ionic states would mean that the mobile electron would jump between ionic states with times lower than $10^{-15}$ seconds (interaction time for the photoabsorption process), giving rise to a relatively high band width that might not be compatible with a tight-binding approximation to the system.

In this paper we report on a detailed XRS study at the Mn K-edge of the $Nd_{0.5}Sr_{0.5}MnO_3$ sample at low temperatures, in the so-called CO phase. A complete phase diagram of the $Nd_{1-x}Sr_xMnO_3$ series has already been described[47-49], interpreting the different phases in terms of CO and OO. In particular, $Nd_{0.5}Sr_{0.5}MnO_3$ shows three phases, an antiferromagnetic CE-type phase below 170 K assigned to COO phase, a ferromagnetic metallic phase stabilised between 170 K and 230 K and a paramagnetic-insulator phase above T = 230 K. Previous x-ray resonant studies performed by Nakamura et al.[30] concluded that the low temperature phase is a COO phase within the classical checkerboard pattern. The present study will shed light on the origin of the observed resonances. We study the azimuthal and polarisation dependence of the (3 0 0), (0 3 0) and (0 5/2 0) reflections at T= 60 K (COO phase). The detailed analysis of the data will show that the checkerboard pattern proposed by Radaelli et al.[9,10] agrees with XRS data, but the simplification to a COO model is not justified. We postulate that the phase transition is better explained as a structural transition differentiating two Mn sites with different oxygen environment.

## 2. EXPERIMENTAL

A single crystal of $Nd_{0.5}Sr_{0.5}MnO_3$ was grown at the Zaragoza University using a floating-zone furnace. The structure corresponds to an orthorhombic distorted perovskite with a=5.515 Å, b=5.452 Å and c=7.552 Å at T ~ 60 K (*Ibmm* setting). A polished $(100)_{cubic}$ surface of a twinned sample was used for the x-ray study. Both (100) and (010) domains were detected. The surface area was about 16 $mm^2$ and the mosaic width (FWHM) was approximately 2 deg. In the following, we will use the Miller indices corresponding to the high-T orthorhombic cell. Then, superlattice (h 0 0) and (0 k 0) attributed to the CO and (0 k/2 0) attributed to the OO (h, k=odd) reflections were analysed.

Figure 2 shows the temperature dependence of the magnetic susceptibility of the single crystal used in the present study in order to check the quality of the sample. The results show two phase transitions at Tc ~ 255 K and at $T_{CO}$ ~ 150 K in agreement with the measurements reported in literature[8,47].

X-ray resonant scattering experiments were performed at the ID20 magnetic scattering ondulator beamline at the European Synchrotron Radiation Facility[50]. The incident beam was monochromatized by a double crystal Si(111) monochromator located between two focusing mirrors. The typical energy resolution of the incident beam at the Mn K-edge was 1 eV with nearly 100% of linear σ polarisation. The sample was mounted with silver paint in a closed cycle refrigerator which could be rotated about the scattering vector to perform azimuthal scans. Polarisation analysis of the scattered beam was performed using a Cu(220) analyser crystal, which gives a scattering angle of 95.9 deg for the energy of Mn K-edge. A schematic view of the experimental configuration together to the definition of the polarisation directions is shown in Figure 3. The linearly polarised σ' and π' components of the scattered beam are perpendicular and parallel to the diffraction plane, respectively. The temperature dependence of the σ-σ' intensity for the (3 0 0) reflection at the Mn K-edge is also shown in Figure 2. The intensity is nearly constant at temperatures below $T_{CO}$ ~ 150 K disappearing above $T_{CO}$ so that it is clearly coupled to the onset of the CE-type antiferromagnetic ordering.

## 3. EXPERIMENTAL RESULTS

Figure 4a shows the intensity versus photon energy of the (3 0 0) reflection across the Mn K-edge for the σ-σ´ scattering channel at different azimuthal angles. The non-zero intensity at energies below the absorption edge shows the existence of Thomson scattering, i.e. structural modulation coming from a small motion of the atoms out of the *Ibmm* symmetry. A broad main resonance is observed at the absorption edge whose intensity strongly depends on the azimuthal angle. The appearing of this resonance at the Mn K absorption edge indicates that the main contribution arises from an energy shift of the absorption edge between different manganese atoms. The strong azimuthal dependence informs us on the anisotropy of the anomalous scattering factors of the Mn atoms, characteristic of ATS reflections. We have also tried to measure the σ-π´ contribution. The observed intensity is smaller than 2% of the σ-σ´ channel, which closely corresponds to the polarisation resolution of the crystal analyser.

Figure 4b shows likewise the intensity of the (0 3 0) reflection as a function of the photon energy together to an absorption spectrum for the sake of comparison. A similar energy dependence and azimuthal behaviour are observed for this reflection compared to the (3 0 0) one. Also in this case, we observed a non-resonant Thomson scattering, a resonant contribution at the edge and finally no σ-π´ contribution was detected. Besides the main resonance, two shoulders at energies around 6565 and 6574 eV were detected for both reflections. The azimuthal dependence at the main resonance of both "odd" reflections shows a minimum for $\phi$=0 deg and a maximum for $\phi$=90 deg, non-vanishing the resonance at the minimum. Moreover, the azimuthal evolution does not only affect the resonance intensities but a small energy shift of the maximum can be observed together with a slight change in the line shape. The peaks of the (3 0 0) and (0 3 0) reflections are located at 6552.6 eV and at 6553.6 eV for $\phi$ =0 deg and $\phi$=90 deg, respectively.

Half-integer reflections (so-called OO reflections) have also been investigated. The (5/2 0 0) reflection was not found either in the σ–σ´ or σ–π´ channels. A strong resonance around the energy of the Mn K-edge (6552 eV) was observed instead for the (0 5/2 0) reflection as it is shown in fig. 5. Scattered intensity is only observed in σ–π' polarisation channel. No Thomson scattering intensity was observed for this half-integer reflection, indicating that it only arises from the anisotropy of the manganese anomalous scattering factor. The gaussian-shaped resonance shows a strong azimuthal dependence disappearing completely for $\phi$=0 deg. Figure 6 shows the azimuthal dependence of the resonance intensities for the three studied reflections. This behaviour nicely fits to a sinusoidal function with $\pi$ period. Reflections (3 0 0) and (0 5/2 0) were also measured at different temperatures in order to check the disappearing of these reflections above the CO transition temperature (see fig. 2).

## 4. ANALYSIS

### *A. Tensorial formalism*

Several structures have been proposed for the low temperature phase of half-doped manganites. In all of them, the crystallographic unit cell is defined as *ax2bxc* with eight Mn atoms relative to the room temperature *Pbnm* (or *Ibmm* in $Nd_{0.5}Sr_{0.5}MnO_3$) structure as it is shown in fig 3. Taking into account that Mn atoms are equivalent along the c-axis, we can operatively reduce to four the number of Mn atoms (denoted as 1,2,3,4 in fig. 3) needed to describe the structure factors of (h 0 0), (0 k 0) and (0 k/2 0) reflections. The observation of non-resonant Thomson scattering for (3 0 0) and (0 3 0) reflections can be originated by the small motion of Mn, Nd(Sr) or O atoms from the high-temperature positions. As a first approximation, we would consider that the Mn atoms do not move and the appearance of non-resonant scattering is ascribed to the small displacement of the O and/or Nd(Sr) atoms. We will show that this approximation is self-consistent with the experimental data.

Within this structural model and using the nomenclature given in fig 3, the structure factors are given by,

$F(h\ 0\ 0) = C_h + f_1 - f_2 + f_3 - f_4$

$F(0\ k\ 0) = C_k + f_1 - f_2 + f_3 - f_4$

$F(0\ k/2\ 0) = f_1 - f_3 + i(f_2 - f_4)$ (1)

Here $C_h$ and $C_k$ denote the Thompson contribution due to the mentioned atomic motion and $f_i$ are the anomalous atomic scattering factors of the i-manganese atoms. We note there is no Thompson contribution for the (0 k/2 0) reflection. The anomalous scattering factor for dipole transitions is a tensor of range 2. Then, it is necessary to determine the three components of each $f_i$ and the spatial orientation (i.e. axis for which the tensor is diagonal). Obviously, it is not possible to determine the whole tensor components with the set of experimental data we have. Therefore, we will discuss the two representative crystallographic models proposed for these manganites: the checkerboard model, where the structure is described by two different Mn sites, one of them with anisotropic local structure (assigned to $Mn^{3+}$) and the other one pseudo-symmetric (assigned to $Mn^{4+}$) and the Zener-polaron model[20] where every Mn atom is locally anisotropic but identical from the electronic point of view. A pictorial view of the Zener polaron model is shown in fig 7. First, we consider the checkerboard model. The anomalous atomic scattering factor (see fig. 3) for the anisotropic atoms (odds ones) in the x´y´z´ axis are given by a diagonal tensor whose components are $f_{||}$ (direction of the anisotropy axis) and $f_{\perp}$ (perpendicular to the anisotropy axis). A diagonal tensor with three identical components $f$ describe the non-anisotropic Mn atoms (even ones)[33]. A model with the anisotropy axis forming 45deg between x and y crystallographic axis has been considered. Then, anomalous atomic scattering tensors in the crystal reference frame are the following:

$$f_1 = 1/2\begin{bmatrix} f_\perp + f_{\parallel} & f_\perp - f_{\parallel} & 0 \\ f_\perp - f_{\parallel} & f_\perp + f_{\parallel} & 0 \\ 0 & 0 & 2f_\perp \end{bmatrix}, \quad f_3 = 1/2\begin{bmatrix} f_\perp + f_{\parallel} & f_{\parallel} - f_\perp & 0 \\ f_{\parallel} - f_\perp & f_\perp + f_{\parallel} & 0 \\ 0 & 0 & 2f_\perp \end{bmatrix}$$

And $f_2 = f_4 = \begin{bmatrix} f & 0 & 0 \\ 0 & f & 0 \\ 0 & 0 & f \end{bmatrix}$ (2)

Using eq. 1, the tensorial structure factors for (h,0,0)/(0,k,0) and (0,k/2,0) reflections are given by equations:

$$F(0,k/2,0) = \begin{bmatrix} 0 & f_\perp - f_{\parallel} & 0 \\ f_\perp - f_{\parallel} & 0 & 0 \\ 0 & 0 & 0 \end{bmatrix},$$

$$F(h,0,0)=F(0,k,0)= \begin{bmatrix} f_\perp + f_{\parallel} - 2f & 0 & 0 \\ 0 & f_\perp + f_{\parallel} - 2f & 0 \\ 0 & 0 & 2f_\perp - 2f \end{bmatrix} \quad (3)$$

On the other hand, in the Zener polaron model the following relations hold: $f_{Z1} = f_{Z4} = f_1$ and $f_{Z3} = f_{Z2} = f_3$ and consequently *F(h,0,0)=F(0,k,0)* = C (non-resonant). We would also like to emphasise that even including a tilt of the anisotropy axis according to a more realistic model, the two “odd” reflections keep on not being resonant. The observation of these resonant reflections discards the Zener polaron model.

Within the checkerboard model the intensity of the resonant reflections considering both σ’ and π’ polarisation components of the scattered x-rays is expressed as follows[33,51]:

$I_{\sigma\sigma'}$ (k/2 0 0) $=I_{\sigma\pi'}$ (k/2 0 0) = 0 (4)

$I_{\sigma\sigma'}$ (h 0 0) $=I_{\sigma\sigma}$’ (0 k 0) = $[C_{h(k)} + 2(f_\perp - f)\cos^2\phi + (f_\perp + f_{\parallel} - 2f)\cos^2\phi]^2 =$

$= [C_{h(k)} + 2(f_\perp - f) + (f_{\parallel} - f_\perp)\sin^2\phi]^2$ (5)

$I_{\sigma\pi'}$ (h 0 0) $=I_{\sigma\pi}$' (0 k 0) = $[(f_{\perp} - f_{||}) \sin\phi \cos\phi \sin\theta]^2$ (6)

$I_{\sigma\sigma}$' (0 k/2 0) = 0 (7)

$I_{\sigma\pi}$' (0 k/2 0) = $[(f_{||} - f_{\perp}) \sin\phi \sin\theta]^2$ (8)

The experimental azimuthal and polarisation dependencies of the studied resonant reflections are nicely reproduced with this model (see fig. 6). Moreover, the respective values of the maximum and minimum azimuthal angles also guarantee that the chosen orientation for the local anisotropy axis in the crystal is right. It is noteworthy that we have not discussed about the origin of the anisotropy yet. In fact, if the anisotropy were assigned to the orientation of a Mn d-orbital ( the anisotropic and isotropic atoms were considered $Mn^{3+}$ and $Mn^{4+}$ ions, respectively) we would obtain the COO model. However, it is important to remark that both types of reflections occur simultaneously and those models trying to describe separately the reflections coming from the “supposed” orbital ordering from those associated to charge ordering are discarded.

*B. The anomalous scattering factor.*

The mechanism of the anomalous scattering is the following: the incoming photon is virtually absorbed to promote a core electron to an empty intermediate empty state leaving a core hole behind. Subsequently, the excited electron decays to the same core hole emitting an outgoing photon with the same energy as the incoming one. For the case of dipolar approximation (the main contribution), the transition goes from a 1s-core state to the empty p-band. The mechanism of x-ray absorption spectroscopy is nearly the same, except for the fact that the electron is really promoted to an empty state of the appropriate symmetry above the Fermi level[52].

The anomalous scattering factor is given by f(E) = f’(E)+ if”(E). The imaginary part is related to the absorption coefficient through the optical theorem by $f''(E)=\left(\frac{mcE}{2e^2h}\right)\mu(E)$, where μ(E) is the absorption coefficient and E is the photon energy. The real part f'(E) is related to f"(E) through the mutual Kramers-Kronig relation.

X-ray Mn-K resonant data have shown that the main resonance appears at energies around the absorption edge. This means that the main differences between the components of the scattering tensor of the involved scattered atoms are placed at the K-edge energy. The energy position of the absorption edge is sensitive both, to the valence state and to the anisotropy of the local environment for different polarisations. In the first case, the difference between the energy edge positions for different valence states is generally referred as the chemical shift. Usually, the chemical shift is determined from x-ray absorption experiments on powder or isotropic samples in such a way that the energy shift corresponds to the non-polarised absorption edge. It measures the difference between the average μ(E) or, the trace of the absorption coefficient tensor within the tensorial framework. In this sense, the chemical shift experimentally found between $Mn^{3+}$ and $Mn^{4+}$ ions is ~ 4.5 eV [44,45] and the appearance of a resonance for the "odd" reflections was qualitatively explained as due to the $Mn^{3+}$ -$Mn^{4+}$ chemical shift[27-31]. In the case of geometrical local anisotropy, the energy position of the absorption edge also depends on the angle between the x-ray polarisation-vector (ε) and the directional axis of anisotropy. This shift that we call *anisotropic shift* is normally smaller than the chemical shift. For instance, a shift between the spectra taken with ε parallel and perpendicular to the *ab* plane of about 2.8 eV is found in $LaSrMnO_4$ manganite[53]. A revision of the anisotropy of XAS up to 1990 is given in the paper of Brouder[43]. In particular, the anisotropic shift in compounds with octahedral coordination has been explained in terms of a tetragonal distortion of the octahedron.

The energy edge position is different for $\mu(E)$ measured with $\varepsilon$ parallel and perpendicular to the tetragonal axis[39,54]. We note that both shifts, chemical and anisotropic, give rise to the so-called derivative effect on the x-ray resonant spectrum although this effect has been only considered as a fingerprint of CO reflections[55].

As a matter of illustration, we have carried out theoretical simulations of the x-ray absorption coefficient $\mu(E)$ at the Mn K-edge using the MXAN program[56]. X-ray absorption spectra were calculated for the two inequivalent Mn positions of the low temperature phase reported by Radaelli et al.[10] after renormalization by the unit cell parameters. Two clusters with 7 ($MnO_6$) and 21 atoms ($MnO_6Nd_8Mn_6$) respectively, have been used for each of the two Mn positions, the Mn-O interatomic distances being d= 1.92 Å for the regular octahedron and $d_{equatorial}$ = 1.92 Å; $d_{axial}$ = 2.06 Å for the tetragonal-distorted octahedron (average distance d =1.967 Å). Figure 8 shows the calculated XANES spectra for the $MnO_6$ and the $MnO_6Nd_8Mn_6$ clusters. The unpolarised XANES spectra of the two Mn geometrical configurations (regular and tetragonal distorted) are shown in the upper part. The effect of a slight homogenous expansion of the lattice, a expanded symmetric cluster (≈2.5%) with average distance, d = 1.967 Å, is also shown for comparison. At the bottom, the two polarised components of the tetragonal-distorted cluster, parallel and perpendicular to the tetragonal axis, are displayed. We observe that the main difference among the several configurations lies in the energy shift of the Mn K-edge. The value of the energy shift is nearly independent of the cluster dimension. We can define the shift arising from the different directions of polarisation as the anisotropic shift ($\delta_{anis}$) and the shift originated by the different average Mn-O distances as the chemical shift ($\delta_{chem}$). The latter can be related to a different charge density on the atoms (valence state). Our calculations give around 1.5 eV and 0.7 eV for $\delta_{ani}$ and $\delta_{chem}$, respectively. We also note that the edge position of the unpolarized spectra does not change when the average Mn-O interatomic distance is alike, independently of the type of distortion (tetragonal or expanded breathing mode).

In conclusion, we have shown that two parameters, $\delta_{anis}$ and $\delta_{chem}$, can account for the anisotropy of the scattering tensor and for the presence of two different kind of Mn atoms, respectively. Now, we will use both parameters to fit the experimental data.

*C. Semi-empirical model.*

The use of theoretical atomic scattering factors to simulate the x-ray scattering signal has been worked reasonably well[33]. As the theoretical simulation was done without considering the valence state of the Mn atoms, it was concluded that the transition was driven by a phonon-softening process giving rise to a differentiation of two kinds of Mn atoms. However, the use of theoretical calculations has some intrinsic limitations. First, the degree of accuracy of the calculated scattering factors. Actually, only a qualitative agreement is fulfilled with the present codes. Secondly, the added difficulty to calculate a spectrum for a system intrinsically inhomogeneous (random distribution of Sr and Nd atoms in $Nd_{0.5}Sr_{0.5}MnO_3$).

Taking these limitations into account, we have chosen a semi-empirical approach to simulate the x-ray resonant scattering data. The spectral line-shape of the XANES spectra (fig. 8) is nearly the same for different polarisations and similar local geometry, the energy position of the K-edge being the main difference between the spectra. We consider that the absorption coefficient (the anomalous scattering factor) has the same spectral line-shape for the two kinds of Mn atoms in the crystal and for the two polarisations. Within this approximation, we define the following relationships between the different components of the anomalous scattering tensors:

$f(E) = f_{anis}(E + \delta_{chem})$ and $f_{\perp}(E) = f_{||}(E+\delta_{ani})$, where f(E) represents the anomalous scattering factor for the symmetric Mn, $f_{||}$ and $f_{\perp}$ are the parallel and the perpendicular components and $f_{anis}$ is defined as the unpolarised anomalous scattering factor $1/3(f_{||}+2 f_{\perp})$ of the tetragonal distorted Mn atom.

We used the experimental unpolarised XANES spectrum of the $Nd_{0.5}Sr_{0.5}MnO_3$ sample at room temperature[45] to obtain the different anomalous scattering tensor components. Figure 9 shows the real (f') and imaginary (f") parts obtained from the experimental x-ray absorption spectrum together with the definition of $f_{||}$, $f_{\perp}$ and f on the basis of the chemical ($\delta_{chem}$) and anisotropic ($\delta_{anis}$) shifts.

We were successful in fitting the experimental x-ray resonant data by using only three variables: $\delta_{ani}$, $\delta_{chem}$ and the independent Thompson scattering term $C_{h(k)}$. As the intensity of the (0 5/2 0) reflection only depends on $\delta_{anis}$ (see eq. 8), we have used this reflection to estimate the value of this parameter. The obtained energy shift between $f_{||}$ and $f_{\perp}$that reproduces the intensity of the (0 5/2 0) reflection is $\delta_{ani} = 1.6\pm0.2$ eV. Fixing this parameter, we have tried to reproduce the x-ray resonant intensities of the "odd" reflections using as fitting parameters $\delta_{chem}$ and $C_{h(k)}$ (see eq. 5). Only a positive isotropic shift ($\delta_{chem} > 0$) was able to fit the spectra and the best value obtained is $\delta_{chem} = 0.7\pm0.1$ eV. The comparison between the experimental and the best-fit model spectra is shown in figure 10. The azimuthal dependence of the intensity for the three reflections is well reproduced. Moreover, this model nicely accounts for the shift in the energy position of the resonances. For example, the energy shift of the main resonant peak for (3 0 0) and (0 3 0) reflections at different $\phi$ is qualitatively reproduced.

## 5. DISCUSSION AND CONCLUSIONS

In summary, x-ray resonant scattering data qualitatively agree with the previous work carried out by Nakamura *et al.*[30] in $Nd_{0.5}Sr_{0.5}MnO_3$. However, our data with higher resolution allows us to give a detailed analysis of the so-called COO low temperature phase of $Nd_{0.5}Sr_{0.5}MnO_3$ using a semi-empirical structural model. This model agrees with the checkerboard pattern proposed for the half-doped manganites in the sense that the structure is described as an ordered distribution of two kinds of Mn atoms from the local symmetry point of view. One of them is considered isotropic, a regular octahedron and the other one shows a strong anisotropy, tetragonal distorted octahedron.

The x-ray scattering data had been presented as experimental evidence for the direct observation of charge and d-orbital ordering in half-doped manganites[27-31]. In this study however, we have described the resonant reflections using a unique tensorial formalism with only three free parameters: the chemical shift ($\delta_{chem}$), the anisotropic splitting ($\delta_{anis}$) and the Thomson scattering contribution. Let us now discuss that the values obtained for $\delta_{chem}$ and $\delta_{anis}$ in $Nd_{0.5}Sr_{0.5}MnO_3$ sample have on the concepts of charge and orbital-ordering, respectively. The obtained $\delta_{chem}$ between the isotropic and anisotropic Mn atoms is 0.7 eV far below from the experimental chemical shift reported[44-46] between $Mn^{3+}$ and $Mn^{4+}$ oxides ( 4.5 eV). Similar chemical shift was determined for $Pr_{0.4}Ca_{0.6}MnO_3$ in another XRS study[57]. This clear difference leads to the conclusion that the assignment of anisotropic or isotropic Mn atom to a 3+ or 4+ valence state is completely unjustified.

On the other hand, a linear relationship between the charge density on an atom and the chemical shift can be considered, as it has been experimentally observed in manganites[44-46]. Following this assumption, the observation of a finite chemical shift has been used for several authors as a proof of charge-segregation. This point must be discussed in detail. First, the energy position of an absorption edge highly depends on the valence state but other effects such as geometry or types of ligands also affect the edge position. Accordingly, differences of about 0.5 eV in the edge position are found for atoms with the same formal valence. Secondly, the interatomic distances also play an important role on the position of the absorption edge. This is probably correlated with the ionic state as we have shown in fig 8. Therefore, it is a bit risky to conclude a real charge disproportionation from a chemical shift of about 0.7 eV.

However, assuming the linear correlation between the chemical shift and the charge state, the charge segregation estimated in $Nd_{0.5}Sr_{0.5}MnO_3$ would be $Mn^{+3.42}$ and $Mn^{+3.58}$. This value agrees with the limit given by XANES spectra of a segregation lower than 0.2 electrons in this system[44,45]. At this point, a full CO ($1e^-$) model can be completely discarded. However, it is noteworthy to discriminate if the electronic state of Mn atoms is either a fluctuating valence state or a pure intermediate valence state. In the case of a fluctuating valence state, $Mn^{3+}$ and $Mn^{4+}$ ions can be temporally distinguished. The diffraction lattice (within the interaction time of the scattering process) should be formed by an ensemble of $Mn^{3+}$ and $Mn^{4+}$ ions with a higher probability to find a $Mn^{3+}$ ion at odd positions and vice-versa at even positions (60%/40% ratio, respectively). This type of ordering scheme would give a resonance with the same azimuthal and energy dependences as for the case of a full ($1e^-$) CO scheme. The only difference would be the intensity of the resonance. It must be around 20% of the intensity for the full CO according to the ordering of only 0.2 $e^-$ instead of 1 $e^-$. In order to check the reliability of the fluctuating model, we compare the experimental data for the (0 3 0) reflection to the calculation obtained from the complete CO model at $\phi$=0 deg (maximum) and $\phi$=90 deg (minimum) as shown in fig.11. It is clearly observed that the $\phi$ dependence of the full CO model is weaker than the experimental. This result is expected because in he limit of zero anisotropy, the dependence on $\phi$ should be costant[35] pure CO reflection (without anisotropy) would not have azimuthal dependence. Accordingly, an intermediate valence state is found for the manganese atoms in $Nd_{0.5}Sr_{0.5}MnO_3$ in agreement with the results obtained by XAS[44-46] for mixed-valence manganites. We remark, that the lack of atomic localisation of $1e^-$ seems to be a general feature of mixed-valence transition-metal oxides as it have also been recently shown in magnetite below the Verwey transition temperature[58].

The main contribution to the resonant scattering in $Nd_{0.5}Sr_{0.5}MnO_3$ is $\delta_{anis}$ whose value is 1.6 eV. Such a value is typical of the splitting between parallel and perpendicular components of a tetragonal-distorted $MnO_6$ octahedron. For instance, such a result is obtained from XANES calculation of the 1s-εp dipole transition at the Mn site in $LaMnO_3$ [39]. Therefore, the anisotropy of the “odd” atoms is originated by the tetragonal distortion of the oxygen environment instead of the Coulomb interaction between the 4p conduction band and the anisotropic 3d atomic orbitals. It is worth reminding that both techniques, XAS and XRS at the K-absorption edges, measure the projected density of p empty states on the absorbing atom. Consequently, the observed anisotropy reflects the anisotropy of these p-states and probably, in correlation with the local geometry, the anisotropy of the locally projected density of d-states should be inferred. In this way, XRS experiments at the Mn $L_{3,2}$ edges on half-doped manganites have reported strong resonances for the so-called OO reflections (in our case half-integer reflection). Unfortunately, these papers interpret the resonant spectra on the basis of a Mn 3d-OO (total or partial)[59-61]. The present paper demonstrate that this interpretation is not supported by XAS and XRS at the K-absorption edge. In summary, the electronic state observed at the Mn atom cannot be considered as the ionic one and consequently, the origin of the anisotropy of the “odd” Mn atoms can not be related to the atomic d-orbital ordering.

The so-called COO phase transition can be explained as a structural phase transition where two non-equivalent crystallographic Mn-sites, tetragonal distorted (odd) and regular (even), are present in the low temperature phase. The CO resonances appear because of the different local geometry of both odd and even Mn atoms. The OO resonance appears from the anisotropy of the anomalous scattering factor of the odd Mn atoms originated by the tetragonal distortion. The difference in the average anomalous scattering factor between both Mn atoms can be interpreted in terms of partial charge segregation arising from the small geometrical differences between them. Obviously, charge segregation is quite common in solid state because charge density on different crystallographic sites is generally different. Amazingly, some authors seem to unify the concept of CO and charge-segregation, the former being a particular case of the latter. However, both concepts, are completely different. In the latter, the electron is shared among several Mn atoms whereas it is localised on one Mn atom in the CO model.

Finally, we want to remark that the non-existence of integer valence states in a very short time scale has strong implications on some widely accepted ideas about the physics of transition metal oxides. For example, the Hubbard –Mott model is based on the assumption that electrons are mainly localised on the atoms and the intra-atomic Coulomb repulsion plays an important role. If electrons are not localised on the atom, the intra-atomic Coulomb repulsion would loss its relevance. Moreover, the single ion Jahn-Teller effect would not operate as the ground state is not the atomic one. However, this conclusion does not mean that strong phonon-electron coupling is still present in these systems.

Summarising, we have shown that XRS and XAS data do not support the ionic description given for half-doped manganites. On the contrary, an itinerant model is necessary to understand the physics of these systems and probably, other related transition-metal oxides.

## ACKNOWLEDGMENTS

We would like to thank ESRF for beam time grating and the ID20 staff, Dr. L. Paolasini, A. Bombardi and N. Kernavanois for their kind assistance in the experiment and M. Benfatto for the MXAN code. This work has been supported by the Spanish CICyT MAT-02-0121 project and DGA.

**REFERENCES.**

1) R. M. Kusters, J. Singleton, D. A. Keen, R. Mc Greevy and W. Hayes, Physica B **155**, 362 (1989)
2) J. M. D. Coey, M. Viret and S. Molnar, Adv. Phys. **48**, 167 (1999)
3) E. Dagotto, T. Hotta and A. Moreo, Phys. Reports **344**, 1 (2001)
4) S.-W- Cheong and H.Y. Hwang. In: Tokura, Y. (Ed.) Contribution to Colossal Magnetoresistence Oxides, Monographs in Condensed Matter Science. Gordon & Breach, London.
5) E. J. W. Verwey, Nature **144**, 327 (1939).
6) J. B. Goodenough, Phys. Rev. **100**, 564 (1955)
7) Y. Tomioka, A. Asamitsu, H. Kuwahara, Y. Moritomo anf Y. Tokura, Phys. Rev. B **53**, R1689 (1996)
8) H. Kuwahara, Y. Tomioka, A. Asamitsu, Y. Moritomo and Y. Tokura, Science **270**, 961 (1995)
9) P. G. Radaelli, M. Marezio, H. Y. Hwang, S.-W. Cheong and B. Batlogg, Phys. Rev. Lett. **75**, 4488 (1995)
10) P. G. Radaelli, D. E. Cox, M. Marezio, S.-W. Cheong, Phys. Rev. B, **55**, 3015 (1997)
11) B. J. Sternlieb, J. P. Hill, U. C. Wildgruber, G. Luke, B. Nachumi, Y. Moritomo, and Y. Tokura, Phys. Rev. Lett. **76**, 2169 (1996)
12) Y. Moritomo, Y. Tomioka, A. Asamitsu, Y. Tokura and Y. Matsui, Phys. Rev. B 51, 3297 (1995)
13) Y. Wakabayashi, Y. Murakami, I. Koyama, T. Kimura, Y. Tokura, Y. Moritomo, K. Hirota,, and Y. Endoh, J. Phys. Soc. Jpn. **69**, 2731 (1999).
14) C. H. Chen, S.-W. Cheong and H. Y. Hwang, J Appl. Phys. **81**, 4326 (1997)
15) S. Mori, C. H. Chen and S.-W. Cheong, Phys. Rev. Lett. **81**, 3972 (1998)
16) J. Blasco, J. García, J. M. De Teresa, M. R. Ibarra, J. Pérez, P.A. Algarabel, C. Marquina and C. Ritter, J Phys.: Cond. Matter **9**, 10321 (1997)
17) S. Larochelle, A. Mehta, N. Kaneko, P. K. Mang, A. F. Panchula, L. Zhou, J. Arthur and M. Greven, Phys. Rev. Lett. **87**, 095502 (2001)
18) M. R. Lees, J. Barrat, G. Balakrishnann, D. McK. Paul, and C. Ritter, Phys. Rev. B **58**, 8694 (1998)

19) Z. Jirak, F. Damay, M. Hervieu, C. Martin, B. Raveau, G. André and F. Boureé, Phys. Rev. B **61**, 1181 (2000)

20) A. Daoud-Aladine, J. Rodríguez-Carvajal, L. Pinsard-Gaudart, M. T. Fernández-Díaz and A. Revcolevschi, Phys. Rev. Lett. **89**, 97205 (2002)

21) H. Aliaga, D. Magnoux, A. Moreo, D. Poilblanc, S. Yunoki, and E. Dagotto, Phys. Rev. B **68**, 104405 (2003)

22) P. Mahadevan, K. Terakura and D. D. Sarma, Phys. Rev. Lett. **87**, 66404 (2001)

23) J. Wang, W. Zhang and D. Y. Xing, J. Phys.:Cond. Matter **14**, 4659 (2002)

24) V. Ferrari, M. D. Towler and P. B. Littlewood, Phys. Rev. Lett. **91**, 227202 (2003)

25) “Resonant Anomalous X-ray Scattering, edited by G. Materlik, C. J. Sparks, and K. Fisher (North-Holland, Amsterdam, 1991).

26) V. E. Dmitrienko, Acta Crystallogr., Sect. A: Found. Crystallogr. **39**, 29 (1983); **40**, 89 (9184); D. H. Templeton and L. K. Templeton, K. Acta Crystallogr., Sect. A: Found. Crystallogr. **41**, 133 (1985), **42**, 478 (1986).

27) Y. Murakami, H. Kawada, H. Kawata, M. Tanaka, T. Arima, Y. Moritomo and Y. Tokura, Phys. Rev. Lett. **80**, 1932 (1998)

28) M. v. Zimmerman, J. P. Hill, Doon Gibbs, M. Blume, D. Casa, B. Keimer, Y. Murakami, Y. Tomioka and Y. Tokura Y, Phys. Rev. Lett. **83**, 4872 (1999).

29) M. v. Zimmerman, J. P. Hill, Doon Gibbs, M. Blume, D. Casa, B. Keimer, Y. Murakami, C. C. Kao, C. Venkataraman, T. Gog, Y. Tomioka, and Y. Tokura, Phys. Rev. B **64**, 195133 (2001)

30) K. Nakamura, T. Arima, A. Nakazawa,Y.Wakabayashi, and Y. Murakami, Phys. Rev. B **60**, 2425 (1999)

31) S. B. Wilkins, P. D. Spencer, T. A. W. Beale, P . D. Hatton, M. v. Zimmermann, S. D. Brown, D. Prabhakaran and A. T. Boothroyd, Phys. Rev. B **67**, 205110 (2003)

32) Y. Murakami, J. P. Hill, D. Gibbs, M. Blume, I. Koyama, M. Tanaka, H. Kawata, H. Arima, Y. Tokura, K. Hirota, and Y. Endoh, Phys. Rev. Lett. **81**, 582 (1998)

33) J. García, M. C. Sánchez, J. Blasco, G. Subías and M. G. Proietti, J. Phys.:Cond. Matter **13**, 3243 (2001)

34) J. Garcia and M. Benfatto, Phys. Rev. Lett. **87**, 189701 (2001)

35) J. Garcia and G. Subías Phys. Rev. B **68**, 127101 (2003)

36) J. García, J. Blasco, M. C. Sánchez, M. G. Proietti and G. Subías, Surf. Rev. and Letters **9**, 821 (2002)

37) M. v. Zimmerman, S. Grenier, C. S. Nelson, J. P. Hill, D. Gibbs, M. Blume, D. Casa, B. Keimer, Y. Murakami, C. C. Kao, C. Venkataraman, T. Gog, Y. Tomioka, and Y. Tokura, Phys. Rev. B **68**, 127102 (2003)

38) S. Ishihara and S. Maekawa, Rep. Prog. Phys**.** **65**, 561 (2002)

39) M. Benfatto, Y. Joly and C. R. Natoli, Phys. Rev. Lett. **83**, 636 (1999)

40) P. Benedetti, J. van den Brink, E. Pavarini, A. Vigliante,and P. Wochner, Phys. Rev. B **63**, 060408 (2001)

41) I. S. Elfimov, V. I. Anisimov, and G. A. Sawatzky, Phys. Rev. Lett. **82**, 4264 (1999)

42) S. di Mateo, T. Chatterji, Y. Joly, A. Stunault, J. A. Paixao, R. Suryanarayanan, G. Dhalenne and A. Revcolevschi, Phys. Rev. B **68**, 024414 (2003)

43) C. Brouder, J. Phys.: Condens. Matter **2**, 701 (1989)

44) G. Subías, J. García, M. G. Proietti and J. Blasco, Phys. Rev. B **56**, 8183 (1997)

45) J. García, M. C. Sánchez, G. Subías and J. Blasco, J. Phys.:Cond. Matter **13**, 3229 (2001)

46) F. Bridges, C. H. Booth, M. Anderson, G. H. Kwei, J.J. Neumeier, J. Snyder, J. Mitchell, J. S. Gardner, and E. Brosha, Phys. Rev. B **63**, 214405 (2001)

47) R. Kajimoto, H. Yoshizawa, H. Kawano, H. Kuwahara, Y. Tokura, K. Ohoyama and M. Ohashi, Phys. Rev. B **60**, 9506 (1999)

48) H. Kawano, R. Kajimoto, H. Yoshizawa, Y. Tomioka, H. Kuwahara and Y. Tokura, Phys. Rev. Lett. **78**, 4253 (1997)

49) H. Kawano-Furukawa, H. Kajimoto, H. Yoshizawa, Y. Tomioka, H. Kuwahara and Y. Tokura, Phys. Rev. B **67**, 174422 (2003)

50) A. Stunault, C. Vettier, F. de Bergevin, N. Bernhoeft, V. Fernández, S. Langridge, E. Lidström, J. E. Lorenzo-Díaz, D. Wermeille, L. Chabert, and R. Chagnon, J. Synchrotron Radiat. **5**, 1010 (1998)

51) Some misprints in the equations of the original article have been corrected.

52) L. B. Sorensen, J. O. Cross, M. Newville, B. Ravel, J. J. Rehr, H. Stragier, C. E. Bouldin, and J. C. Woicik in "Resonant Anomalous X-ray Scattering, edited by G. Materlik, C. J. Sparks, and K. Fisher (North-Holland, Amsterdam, 1991) pag 389.

53) We have recently measured the polarised Mn-K XANES spectra of $La_{1-x}Sr_{1+x}MnO_4$ samples. To be published

54) J. García, M. Benfatto, C. R. Natoli, A. Bianconi, A. Fontaine and H. Tolentino, J. Chem. Phys. **132**, 295 (1989)

55) S. Grenier, A. Toader, J. E. Lorenzo, Y. Joly, B. Grenier, S. Ravy, L. P. Regnault, H. Renevier, J. Y. Henry, J. Jegoudez, and A. Revcolevski, Phys. Rev. B **65**, 180101 (2002)

56) Benfatto, S. della Longa and C. R. Natoli, J. Synchrotron Rad. **10**, 51 (2003)

57) Grenier, J. P. Hill, D. Gibbs, K. J. Thomas, M v. Zimmerman, C. J. Nelson, V. Kiryukhin, Y. Tokura, Y. Tomioka, D. Casa, T. Gog and C. Venkataraman, Phys. Rev. B **69**, 134419 (2001)

58) J. García, G. Subías, M. G. Proietti, J. Blasco, H. Renevier, Y. Joly, J. L. Hodeau, and Y. Joly, Phys. Rev. B **63**, 054110 (2001)

59) S. B. Wilkins, P. D. Hatton, M. D. Roper, D. Prabhakaran and A. T. Boothroyd, Phys. Rev. Lett. **90**, 187201 (2003)

60) S. B. Wilkins, P. D. Spencer, P. D. Hatton, S. P. Collins, M. D. Roper, D. Prabhakaran and A. T. Boothroyd, Phys. Rev. Lett. **91**, 167205 (2003)

61) K. J. Thomas, J. P. Hill, Y-J. Kim, S. Grenier, P. Abbamonte, L. Venema, A. Rusydi, Y. Tomioka, Y. Tokura, D. F. McMorrow and M. V. Veenendaal, unpublished.

## Figure captions:

Fig. 1. Scheme of the CE-type charge-orbital ordering model (denoted as chekerboard) of half-doped manganites. Elongated lobules represent the occupied $e_g$ orbital of the $Mn^{3+}$ ions and closed circles represent the $Mn^{4+}$ ions. The zig-zag pattern is marked by dotted lines.

Fig. 2. Magnetic susceptibility of the $Nd_{0.5}Sr_{0.5}MnO_3$ sample as a function of temperature (continuous line). The temperature dependence of the intensity of the (3 0 0) σ-σ' reflection is also shown as fingerprint of the onset of the charge-orbital phase transition (circles).

Fig. 3. Unit cell of the low temperature phase of $Nd_{0.5}Sr_{0.5}MnO_3$ together with the diffraction configuration for the x-ray scattering experiment. Elongated octahedra indicate the geometrical anisotropic "odd" atoms while "even" ones are represented by regular octahedra. The two studied planes, corresponding to (h 0 0) and (0 k 0) reflections are also indicated. The azimuthal angles, ϕ and ϕ' are the rotation angles around the respective diffraction **Q** vector. σ–σ' and σ–π' polarization geometries are also indicated for the two planes. The axes for which the atomic anomalous scattering tensor is diagonal are shown in the figures at the right.

Fig. 4. Intensity of the (3 0 0) and (0 3 0) reflections as a function of the energy at different azimuthal angles in the σ–σ' channel. Panel (a) shows data for the (300) reflection and panel (b) for the (0 3 0) one. The reference for the azimuthal angle (ϕ,ϕ' = 0) corresponds to the crystallographic directions [0 1 0] for (3 0 0) and [1 0 0] for (0 3

0) reflections, respectively. The fluorescence spectrum is also shown for comparison (thick solid line).

Fig. 5. Intensity of the (0 5/2 0) reflection versus energy at different azimuthal angles in the σ−π’ channel. The same criteria as for the (0 3 0) reflection was taken for the zero of the azimuthal angle.

Fig. 6. Azimuthal behavior of the scattered intensity at the resonance for the three studied reflections: (3 0 0) σ−σ’ channel (gray circles); (0 3 0) σ−σ’ channel (open circles) and ( 0 5/2 0) σ−π’ channel (squares).

Fig 7. Pictorial view of the Zener polaron model. $Mn_1$-$Mn_2$ pairs are marked by dashed lobules.

Fig. 8. Theoretical MXAN calculation of the XANES spectra of $MnO_6$ and $MnO_6Nd_8Mn_6$ clusters. The upper curves show the unpolarized spectra for the two crystallographic Mn positions: tetragonal distorted (continuous line) and a symmetric octahedron (dashed lines). The spectra for a symmetric octrahedron with a breathing mode distortion (2.5% expanded) is also shown for comparison (dotted line). Lower curves show the parallel (circles) and perpendicular (triangles) components to the tetragonal axis of the tetragonally distorted cluster.

Fig 9. Real f’ (lower curves) and imaginary f” (upper curves) parts of the different tensor components f (dashed line), $f_{\perp}$ (dotted line) and $f_{||}$ (continuous line) of the Mn anomalous scattering factor obtained from the experimental XANES spectra of $Nd_{0.5}Sr_{0.5}MnO_3$. The definition of $\delta = f - f_\perp$ and $\delta_{anis} = f_\perp - f_{||}$, being $\delta=\delta_{chem}-1/3\delta_{anis}$ are indicated in the inset.

Fig 10. X-ray resonant scattering data at the Mn K-edge for the (3 0 0), (0 3 0) and (0 5/2 0) reflections compared to the semi-empirical best-fit model. The resulting parameters used are $\delta_{ani}= 1.6\pm0.2$ eV, $\delta_{chem} = 0.7\pm0.1$ eV and the Thomson scattering factors $C_{h,k}$=0.43, 0.35 and 0 for (300), (030) and (05/20) reflections, respectively. Experimental maximum ($\phi$=90 deg ) and minimum curves ($\phi$= 0 deg) data are represented by symbols while lines correspond to the best-fit model.

Fig 11. Theoretical simulation for the fluctuating charge-ordering model ($\delta_{chem}$ = 4.5 eV) at $\phi$ = 90 deg (maximum, solid line) and $\phi$= 0 deg (minimum, dashed line) azimuthal angles, compared to the experimental data of the “odd” reflections ($\phi$ = 90 deg, squares and $\phi$ =0 deg, circles).

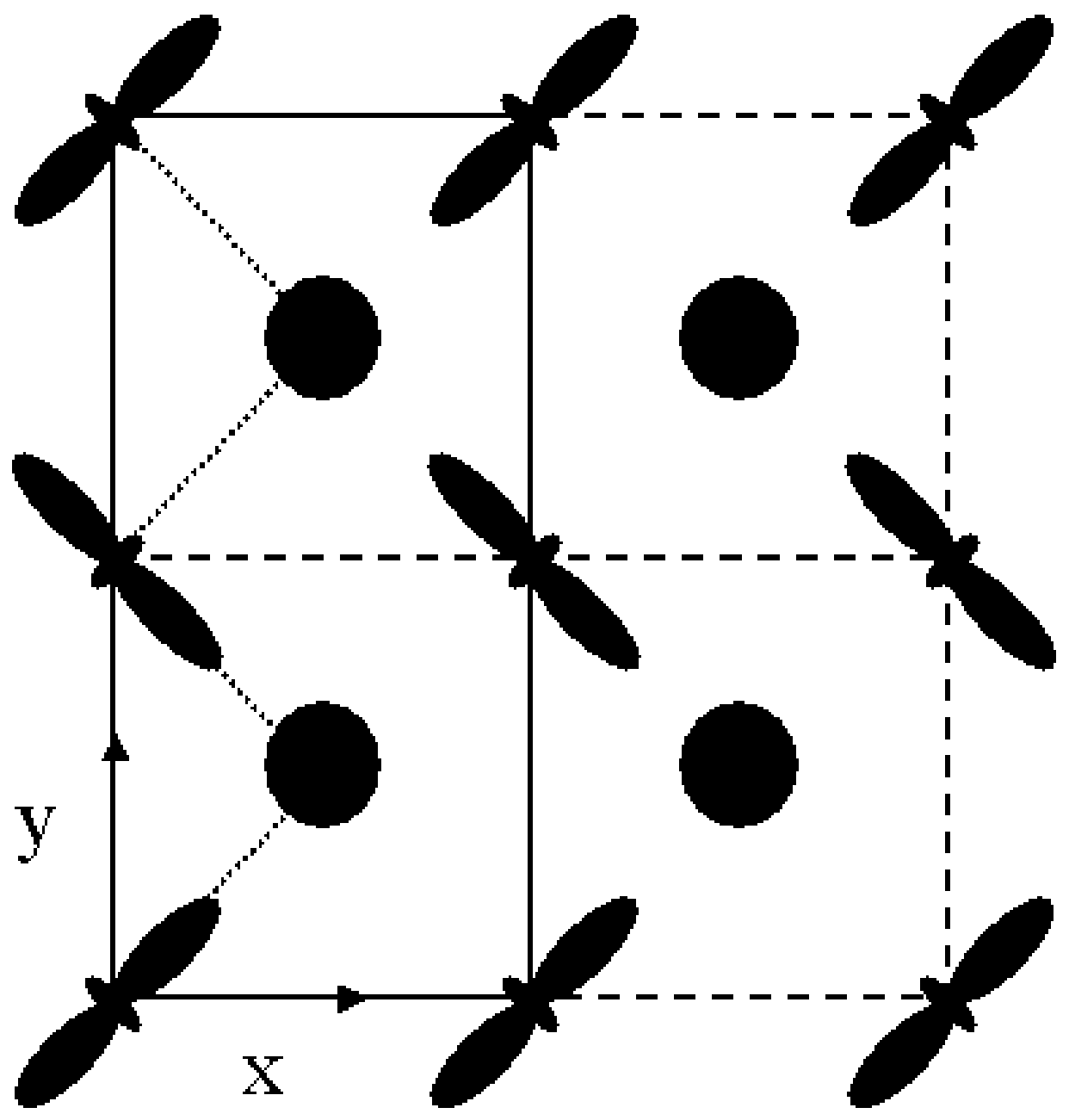
y
x

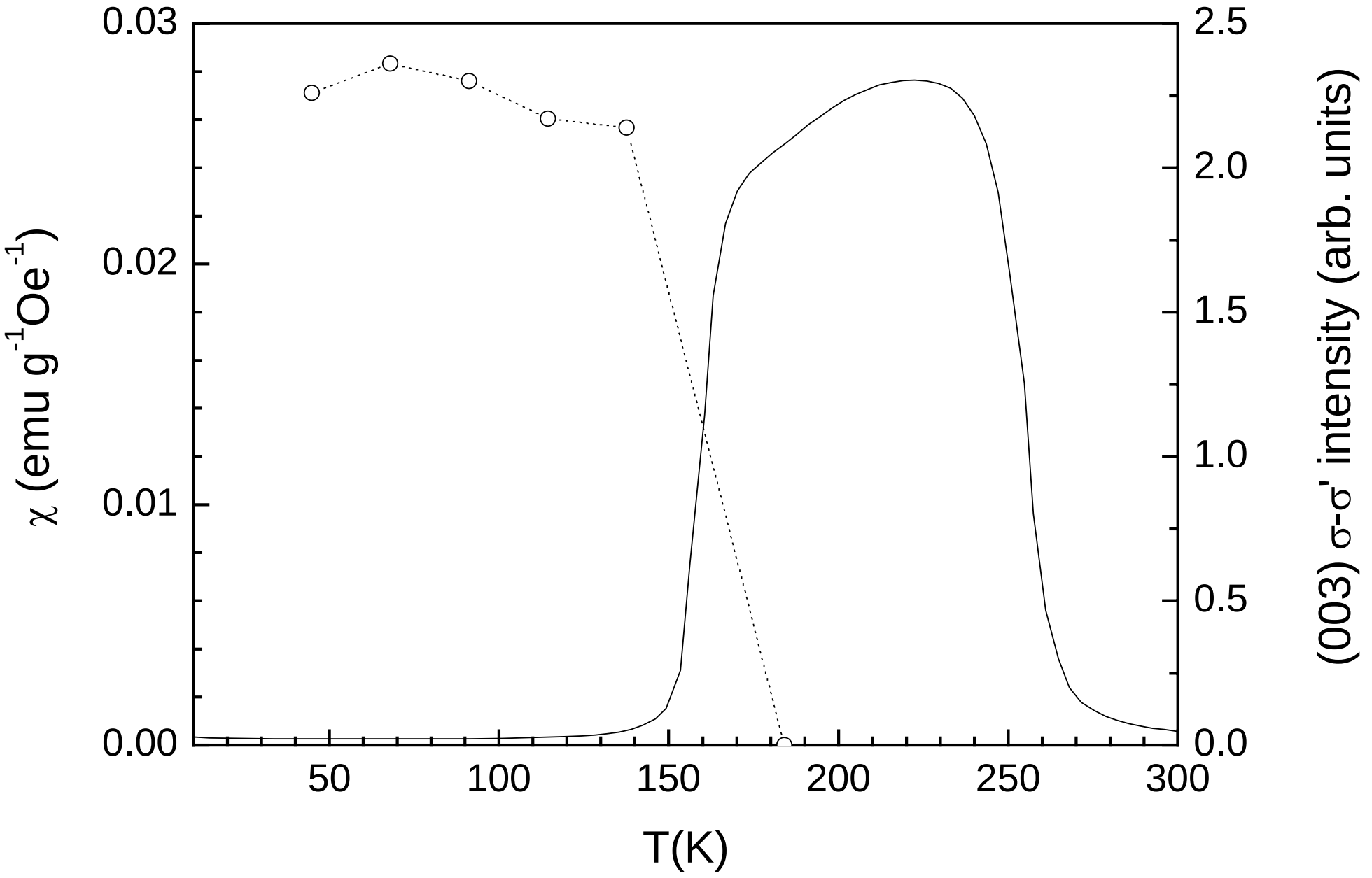

*J. Herrero-Martin et al.* *Fig.3*

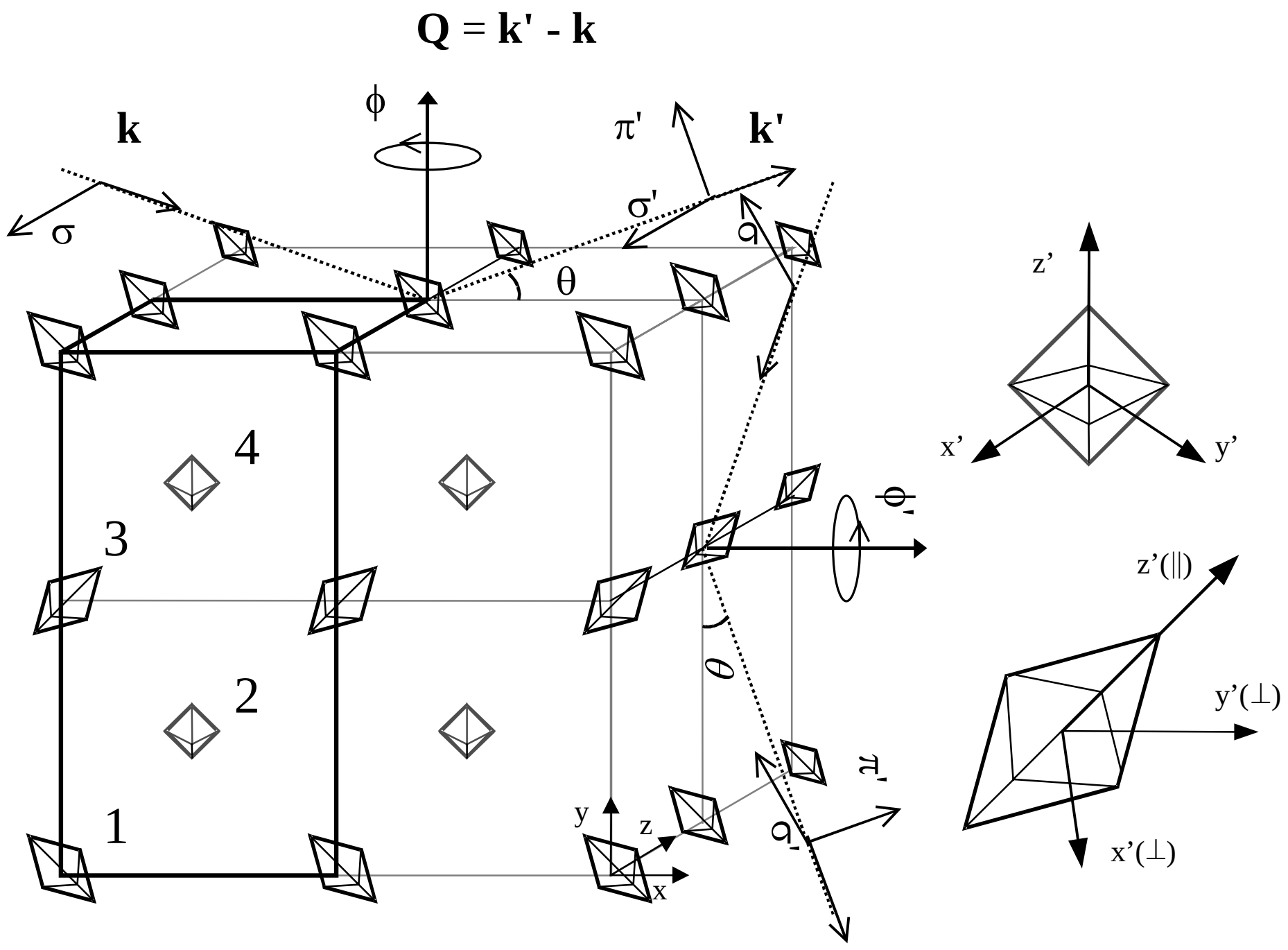

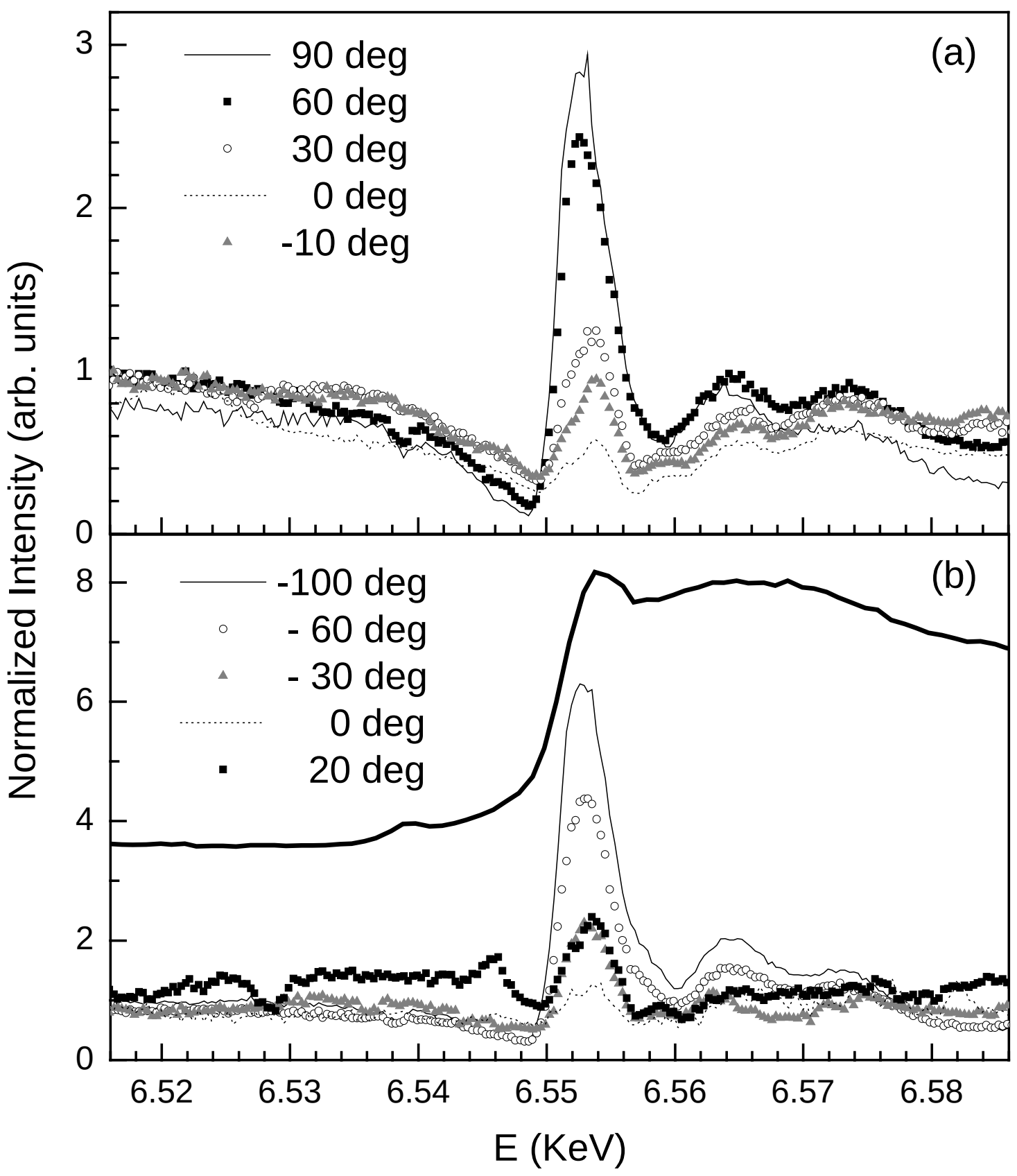

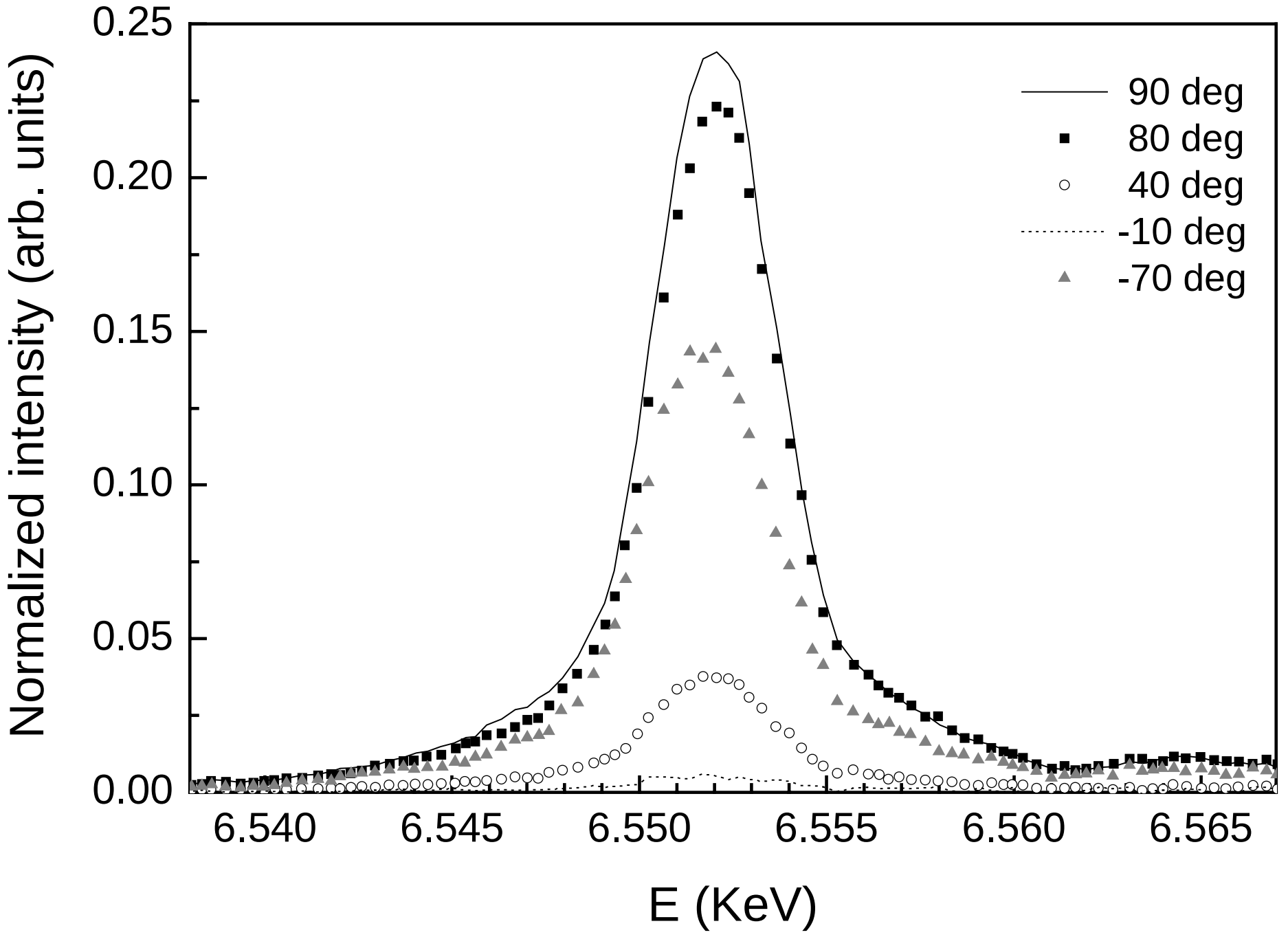

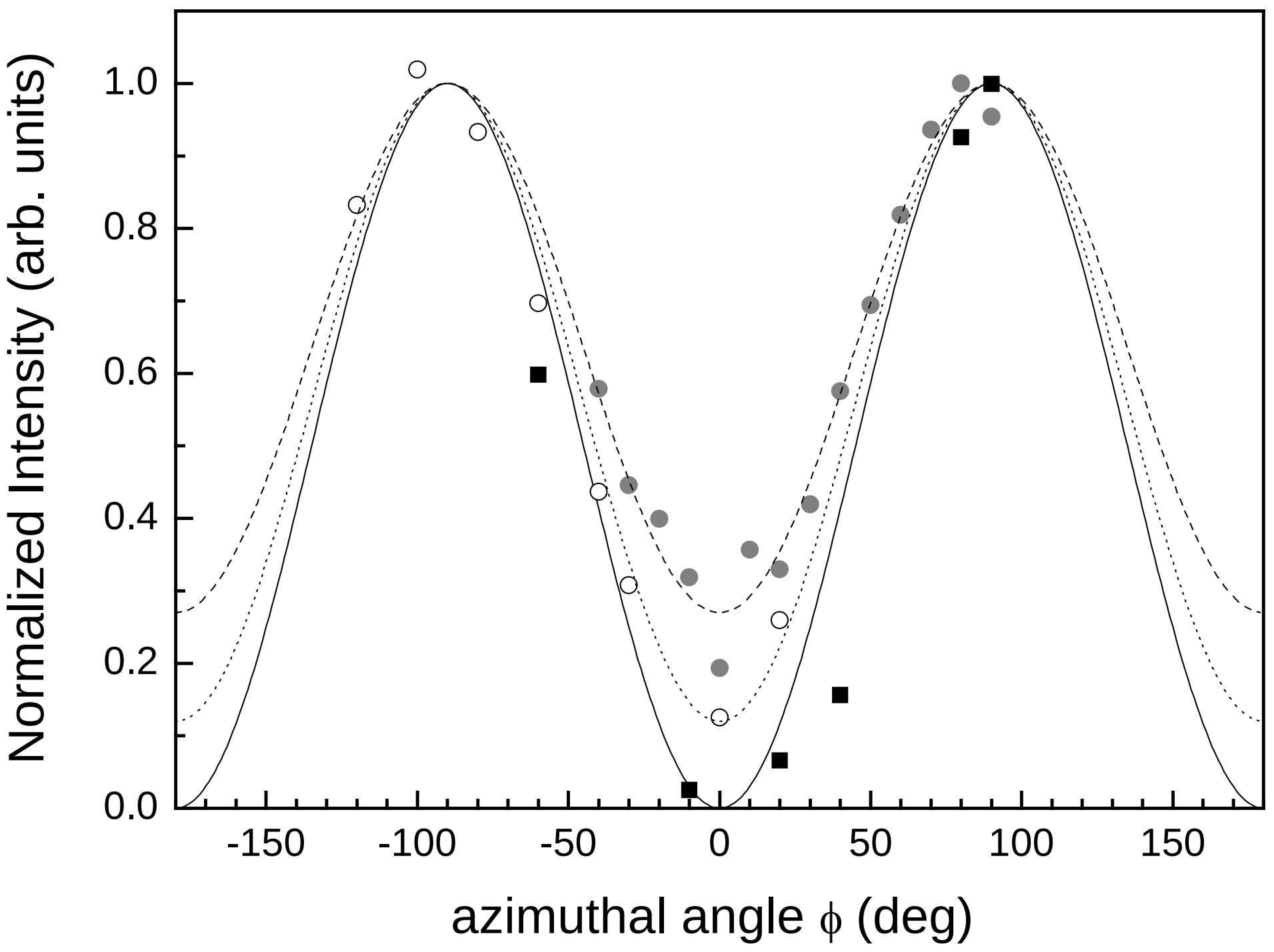

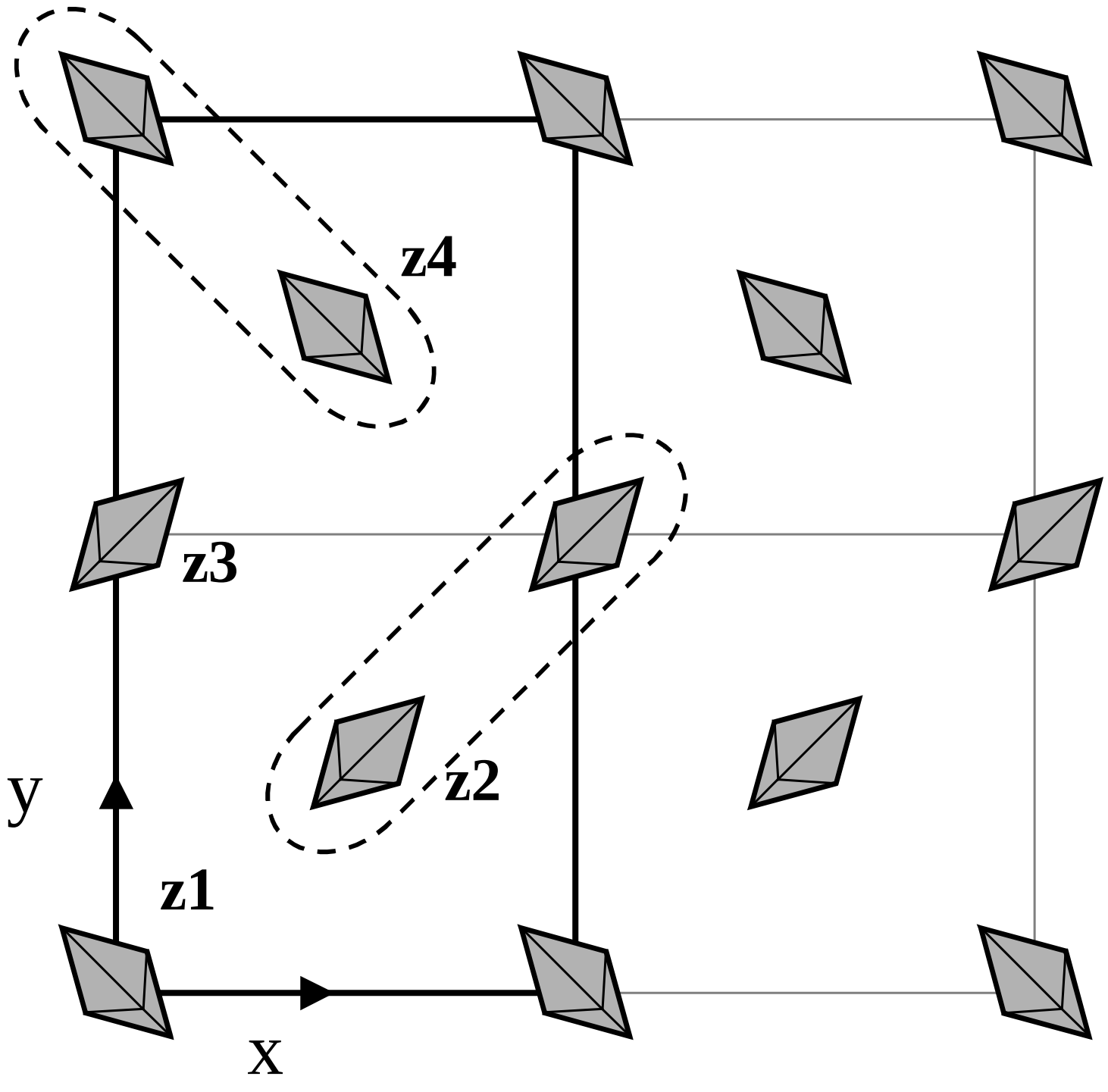

**J. Herrero-Martin et al.** **Fig.8**

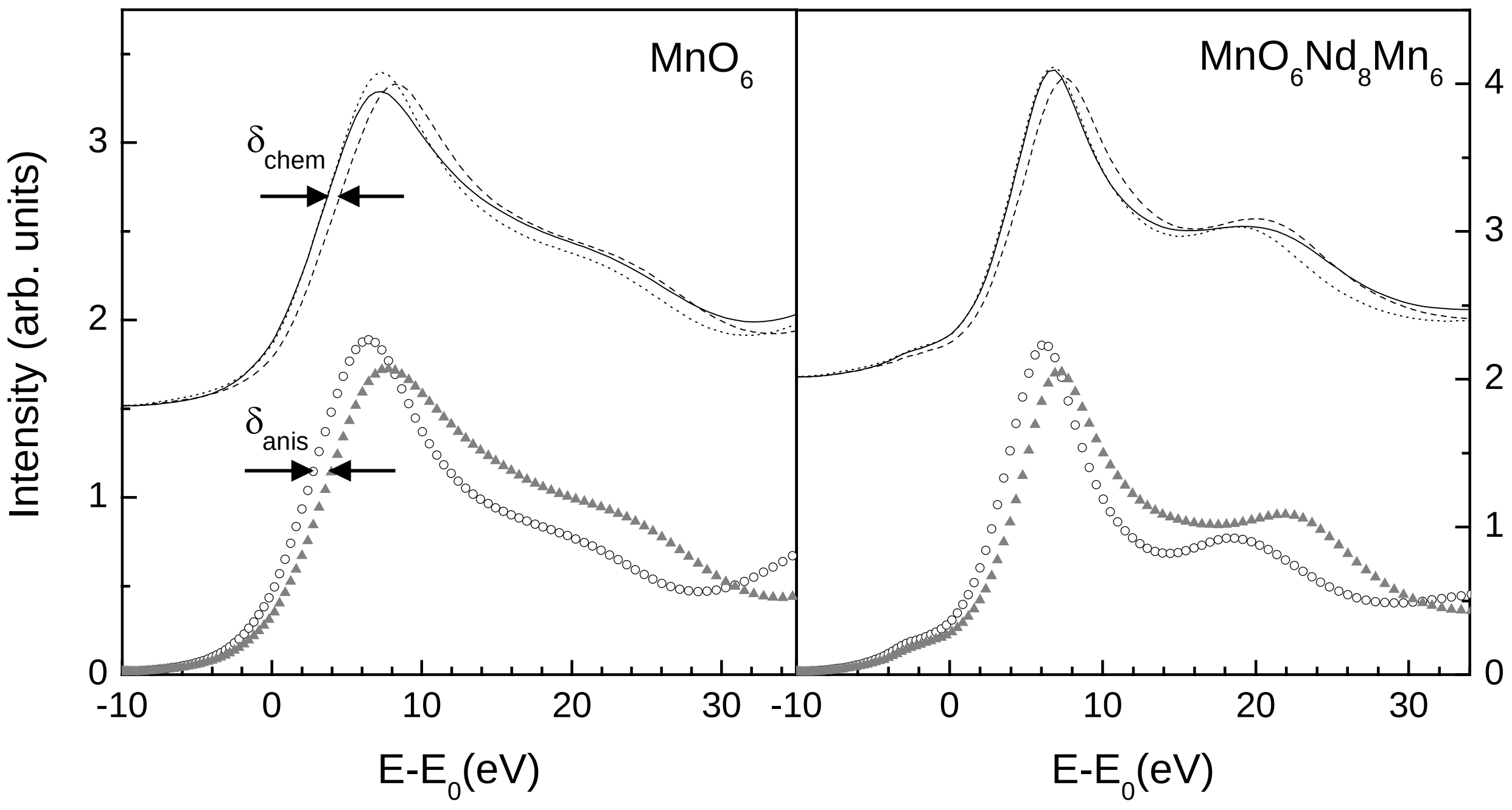

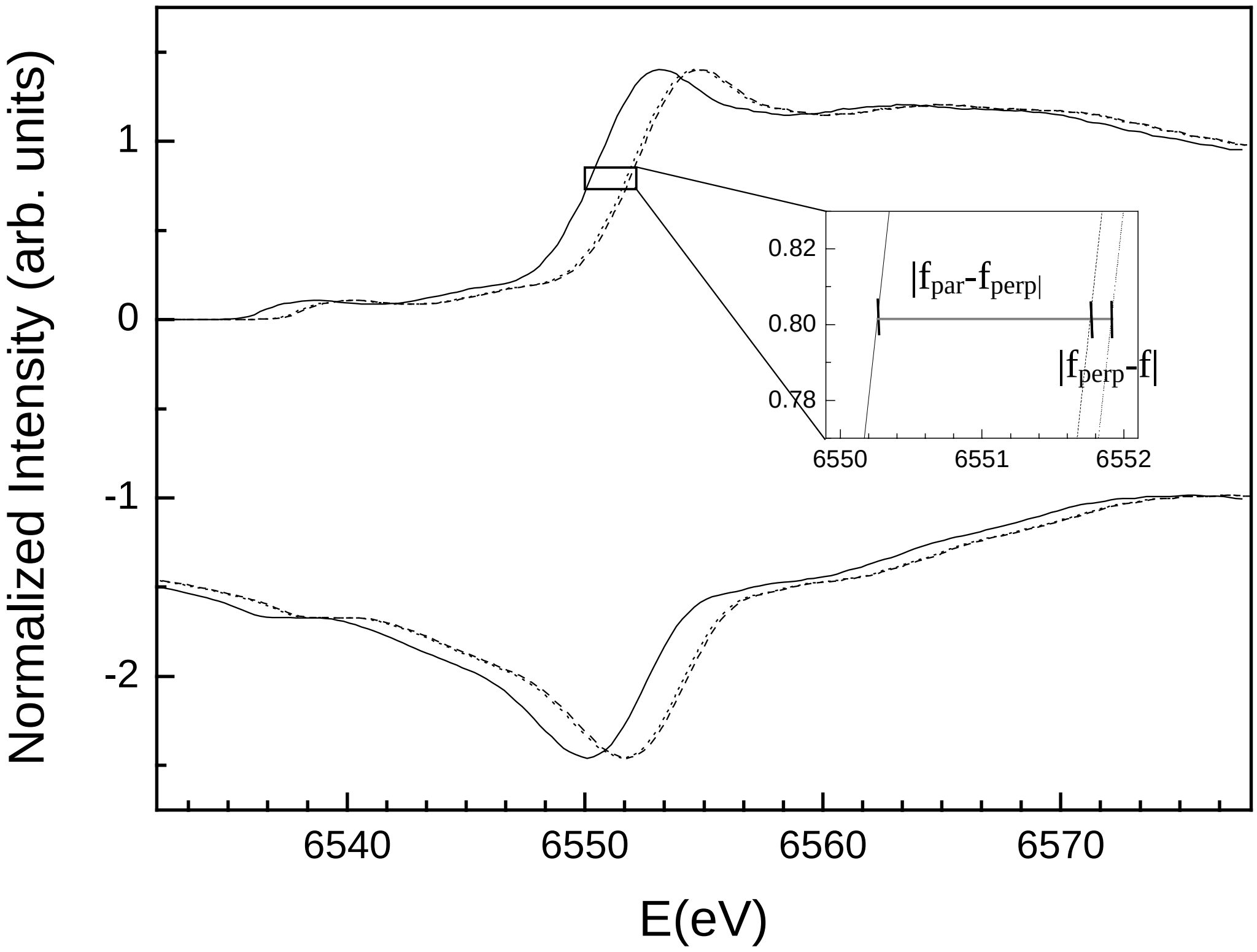
Normalized Intensity (arb. units)
1
0
-1
-2
6540
6550
6560
6570
E(eV)
0.82
0.80
0.78
6550
6551
6552
|fpar-fperp|
|fperp-f|

**J. Herrero-Martin et al. Fig.10**

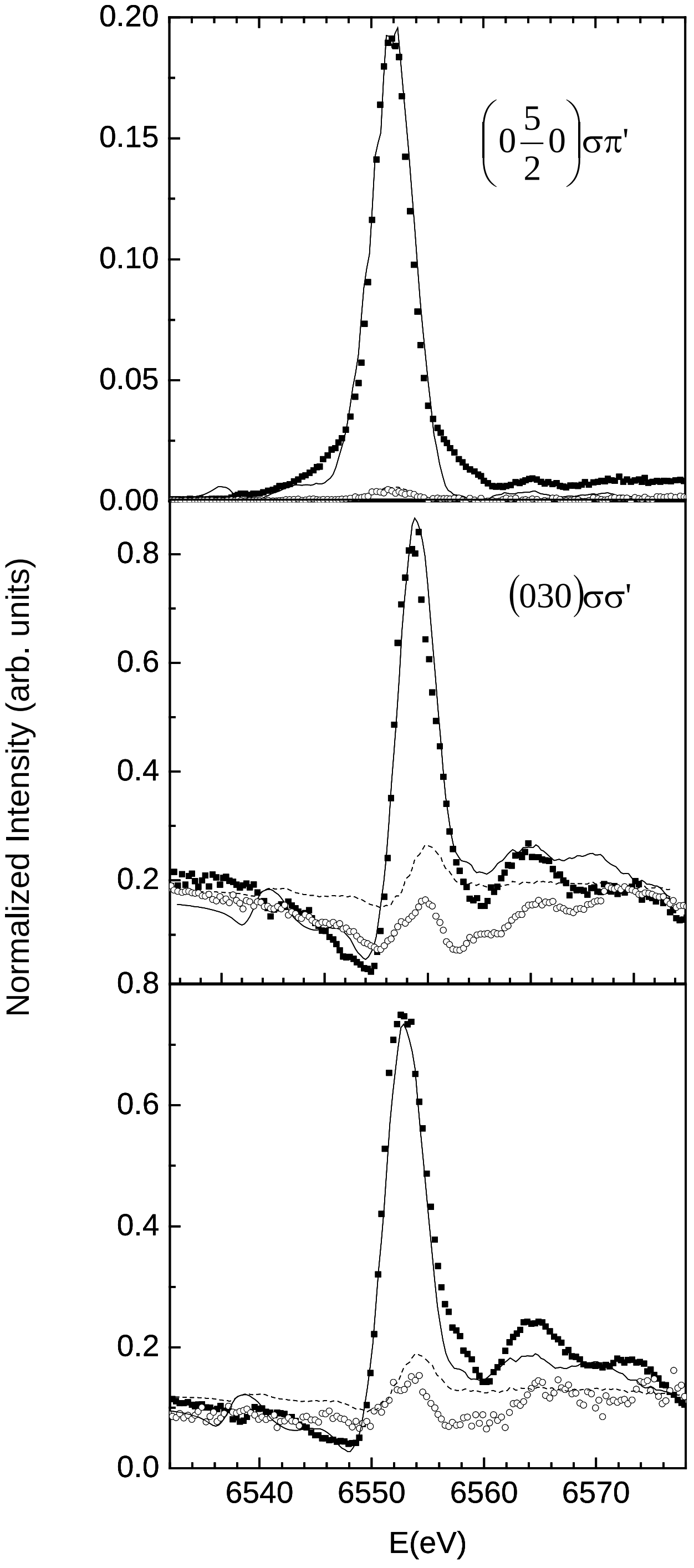

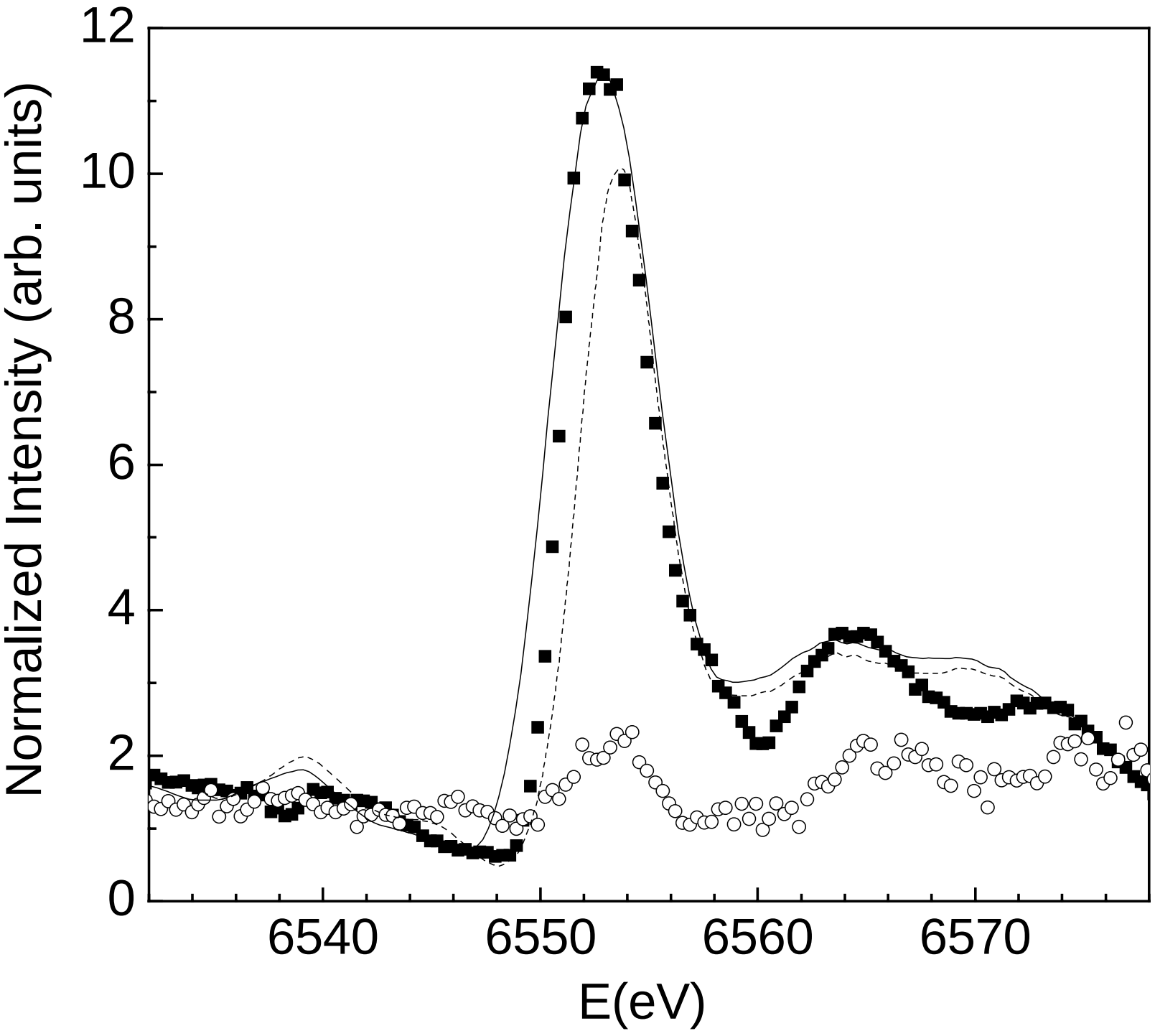